%% file: conference.tex
\documentclass[conference]{IEEEtran}
\ifCLASSINFOpdf

\else

\fi

\usepackage{hyperref}

\usepackage{cite}
\usepackage[cmex10]{amsmath}
\usepackage{amsmath}
\usepackage{algorithm}
\usepackage{algorithmic}
\usepackage{stfloats}
\usepackage{multicol,multienum}
\usepackage{multirow}
\usepackage[dvips]{graphicx}
\usepackage{wrapfig}
\usepackage{color}

\usepackage{cite}
\usepackage[cmex10]{amsmath}
\usepackage{amsmath}
\usepackage{algorithm}
\usepackage{algorithmic}
\usepackage{stfloats}
\usepackage{multicol,multienum}
\usepackage{multirow}
\usepackage[dvips]{graphicx}

\usepackage[tight,footnotesize]{subfigure}
\usepackage{wrapfig}
\usepackage{color}

 \usepackage{booktabs}
 \usepackage{multirow}

\usepackage{mathtools}

\usepackage{float}
\usepackage{bm}

\usepackage{pifont}

\usepackage{bbding}
\usepackage{bbding}
\usepackage{pifont}
\usepackage{wasysym}
\usepackage{amssymb}
\usepackage{amsfonts}

\IEEEoverridecommandlockouts

\begin{document}

\title{\huge Stackelberg Game–Driven Defense for ISAC Against Channel Attacks in Low-Altitude Networks}

\author{
    \IEEEauthorblockN{
    Jiacheng Wang$^{1}$, 
    Changyuan Zhao$^{1}$,
    Dusit Niyato$^{1}$,
    Geng Sun$^{2}$,
    Weijie Yuan$^{3}$,
    Abbas Jamalipour$^{4}$,
    Tao Xiang$^{5}$
    }   
    \IEEEauthorblockA{
$^{1}$College of Computing and Data Science, Nanyang Technological University, Singapore\\
$^{2}$College of Computer Science and Technology, Jilin University, Changchun, China\\ 
$^{3}$School of System Design and
Intelligent Manufacturing, Southern University of Science and Technology, Shenzhen, China\\
$^{4}$School of Electrical and Computer Engineering,
University of Sydney, Sydney, Australia\\
$^{5}$College of Computer Science, Chongqing University, Chongqing, China
}
}

\maketitle

\begin{abstract}
The increasing saturation of terrestrial resources has driven economic activities into low-altitude airspace. These activities, such as air taxis, rely on low-altitude wireless networks, and one key enabling technology is integrated sensing and communication (ISAC). However, in low-altitude airspace, ISAC is vulnerable to channel-access attacks, thereby degrading performance and threatening safety. To address this, we propose a defense framework based on a Stackelberg game. Specifically, we first model the system under attack, deriving metrics for the communication and the sensing to quantify performance. Then, we formulate the interaction as a three-player game where a malicious attacker acts as the leader, while the legitimate drone and ground base station act as followers. Using a backward induction algorithm, we obtain the Stackelberg equilibrium, allowing the defenders to dynamically adjust their strategies to mitigate the attack. Simulation results verify that the proposed algorithm converges to a stable solution and outperforms existing baselines, ensuring reliable ISAC performance for critical low-altitude applications.

\end{abstract}

\begin{IEEEkeywords}
Low-altitude wireless networks (LAWNs), integrated sensing and communication (ISAC), channel access attack, Stackelberg game.
\end{IEEEkeywords}
\vspace{-0.35cm}

\IEEEpeerreviewmaketitle

\input{sections/section1}
\input{sections/section2}

\input{sections/section3}
\input{sections/section4}
\input{sections/section5}

\section{Conclusion}\label{CC}

This paper addresses the critical challenge of securing the ISAC performance in LAWNs against channel-access attacks on time-sensitive targets. We have proposed a defense framework based on a Stackelberg game, where we model the attacker as the leader and the legitimate drone with RIS and an ISAC BS as followers. 
Through the proposed BI algorithm, the system can achieve a stable equilibrium, thereby effectively improving both the communication SINR and the sensing AoI under attacks. Simulation results have verified the method's effectiveness and superiority, supporting the construction of reliable LAWNs. Future work will extend this framework by considering additional constraints, such as energy consumption and flight control, to further enhance its practical applicability.

\bibliography{ref}{}
\bibliographystyle{IEEEtran}

\end{document}

%% file: sections/section1.tex
\section{Introduction}

As ground resources become increasingly constrained, economic activities are gradually expanding into the low-altitude airspace, giving rise to applications such as drone-based logistics and urban air taxis~\cite{zhao2025generative}. To coordinate these operations, low-altitude wireless networks (LAWNs) have emerged as the crucial infrastructure that integrates communication, computing, and control capabilities for aerial vehicles. Among the key enablers of LAWNs, integrated sensing and communication (ISAC) plays a vital role, which allows the same waveform to support both data transmission and environmental perception~\cite{wang2024generative}. This dual-function nature enables a BS to exchange information with drones while detecting obstacles or tracking time-sensitive targets. Furthermore, drones with reconfigurable intelligent surfaces (RIS) can further reflect and shape signals, extending coverage and improving link reliability~\cite{jiang2025integrated}.

However, the openness of the low-altitude airspace exposes LAWNs to malicious interference. Particularly, channel-access attacks can jam control channels or inject noise, delaying updates and threatening air-traffic safety~\cite{wei2020classification}. So far, several works have been proposed to handle jamming, such as Bayesian or non-cooperative Stackelberg games that optimize transmit power and precoding under attack~\cite{liu2024game}. Nevertheless, these defenses assume static or independent attackers, overlooking their adaptive behavior, which limits the defensive performance in practical LAWNs.

To address these challenges, we propose a Stackelberg game-based defense framework to enhance ISAC performance against adaptive channel-access attacks in LAWNs. In this hierarchical game, the attacker acts as the leader, launching the attack first, while the legitimate RIS-assisted drone and the ISAC BS serve as followers, adjusting their sensing and transmission strategies accordingly. This formulation reflects real-world dynamics where LAWNs must respond to evolving threats and optimize jointly for communication reliability and sensing freshness. In general, the main contributions are summarized as follows.
\begin{itemize}
\item We introduce an age of information (AoI) metric to quantify the freshness of sensing data, capturing temporal dynamics essential for time-sensitive targets in LAWNs.

\item We formulate an ISAC performance optimization problem as a three-player Stackelberg game among the attacker, the RIS-assisted drone, and the BS, characterizing their hierarchical interaction and dynamic strategies.

\item We develop a backward-induction (BI) algorithm to derive the Stackelberg equilibrium and verify through simulations that the iterative process converges to stable strategies for all players.

\end{itemize}

%% file: sections/section2.tex
\section{System model}\label{SM}

\subsection{Low-Altitude Economy Network System Overview}
\begin{figure}[t]
  \centering
  \includegraphics[width=7cm]{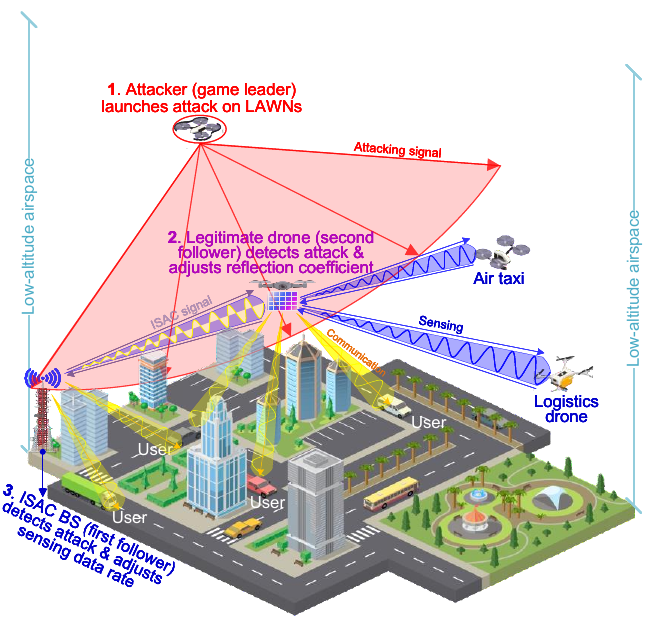}\\
  \caption{System model of the LAWNs under channel access attack. Here, the malicious UAV first launches an attack by injecting additional noise to the wireless channel. In response, the legitimate UAV with RIS and BS respectively adjust the gain and sensing data generation rate to mitigate the impact of the attack, thereby maintaining the ISAC performance.}
  \vspace{-0.3cm}
  \label{SYSTEMMODEL}
\end{figure}

Fig.~\ref{SYSTEMMODEL} illustrates the considered LAWNs scenario comprising a dual-function ISAC BS, multiple mobile users, a legitimate drone equipped with an RIS, time-sensitive targets, and a malicious drone that attempts channel-access attacks. The ISAC BS, equipped with $M$ transmit antennas, simultaneously communicates with users and drones, emitting sensing waveforms to monitor nearby targets. The legitimate RIS-assisted drone relays both communication and sensing services to extend BS coverage. Meanwhile, a malicious drone performs a channel access attack and subsequently injects additional noise into relevant links to degrade the ISAC performance.

\subsection{ISAC Signal Model at BS}

Let $M$ be the number of BS transmit antennas, $N$ the number of users, ${\bf s}_{\rm r}\!\in\!\mathbb{C}^{M\times 1}$ the sensing signal, and ${\bf s}_{\rm c}\!\in\!\mathbb{C}^{N\times 1}$ the communication symbols.
The sensing and communication beamformers are denoted as ${\bf B}_{\rm r}\in\mathbb{C}^{M\times M}$ and ${\bf B}_{\rm c}\in\mathbb{C}^{M\times N}$, respectively. 
The transmitted signal from the BS is
\begin{equation}\label{eq1}
{\bf s} = {\bf B}_{\rm r}{\bf s}_{\rm r} + {\bf B}_{\rm c}{\bf s}_{\rm c} \;=\; {\bf B}\,\hat{\bf s},
\end{equation}
where ${\bf B}=\big[\,{\bf B}_{\rm r},{\bf B}_{\rm c}\,\big]\in\mathbb{C}^{M\times(M+N)}$ is the ISAC transmission beamforming matrix,
and $\hat{\bf s}=\big[\,{\bf s}_{\rm r}^{\rm T},{\bf s}_{\rm c}^{\rm T}\,\big]^{\rm T}\in\mathbb{C}^{(M+N)\times 1}$ is the stacked signal vector.
We adopt a pseudo-random sensing waveform satisfying $\mathbb{E}[{\bf s}_{\rm r}]={\bf 0}$ and $\mathbb{E}[{\bf s}_{\rm r}{\bf s}_{\rm r}^{H}]={\bf I}_{M}$, 
while the data symbols follow ${\bf s}_{\rm c}\!\sim\!\mathcal{CN}({\bf 0},{\bf I}_{N})$ and are independent of ${\bf s}_{\rm r}$~\cite{liu2020joint}, 
where ${\bf I}$ is the identity matrix.
Therefore, the covariance of the transmitted signal is
\begin{equation}\label{eq2}
{\bf R}
= \mathbb{E}[{\bf s}{\bf s}^{H}]
= {\bf B}{\bf B}^{H}
= {\bf B}_{\rm r}{\bf B}_{\rm r}^{H} + \sum_{i=1}^{N} {\bf R}_{i},
\end{equation}
where ${\bf R}_{i}={\bf b}_{{\rm c},i}{\bf b}_{{\rm c},i}^{H}$ is the rank-one contribution of the $i$-th user’s beam ${\bf b}_{{\rm c},i}$ in ${\bf B}_{\rm c}=[{\bf b}_{{\rm c},1},\ldots,{\bf b}_{{\rm c},N}]$. 

\subsection{Wireless Channel Model}

Let the legitimate drone-mounted RIS have $P$ reflecting elements. We denote BS--drone, drone--user~$i$, and BS--user~$i$ channels as ${\bf H}\!\in\!\mathbb{C}^{P\times M}$, ${\bf h}_{1,i}\!\in\!\mathbb{C}^{P\times 1}$, and ${\bf h}_{2,i}\!\in\!\mathbb{C}^{M\times 1}$, respectively. We consider a block-fading model, where channels remain quasi-static within each frame, rendering Doppler effects negligible at typical low-altitude UAV speeds.
As users are typically surrounded by rich scatterers, we model ${\bf h}_{1,i}$ and ${\bf h}_{2,i}$ as Rayleigh fading channels, with large-scale path loss embedded in their variances.
Given the presence of the LoS component, the BS–drone link is modeled as
\begin{equation}
\label{eq:rician_block}
\left\{
\begin{aligned}
{\bf H} \;&=\; {\bf H}_{\rm LoS} \,+\, {\bf H}_{\rm NLoS}, \\[2pt]
{\bf H}_{\rm NLoS} \;&\sim\; \mathcal{CN}\!\big({\bf 0},\, \Sigma_{\rm RIS}\!\otimes\!\Sigma_{\rm BS}\big), \\[2pt]
{\bf H}_{\rm LoS} \;&=\; {\bf a}_{2}(\theta_2)\,{\bf a}_{1}^{H}(\theta_1).
\end{aligned}
\right.
\end{equation}
Here, $\Sigma_{\rm BS}\geq{\bf 0}$ and $\Sigma_{\rm RIS}\geq{\bf 0}$ are Hermitian positive semidefinite spatial correlation matrices with unit diagonal entries, $\otimes$ is the Kronecker product, and $\theta_1$ and $\theta_2$ denote the AoD at the BS and AoA at the drone, respectively.

The steering vectors ${\bf a}_{1}(\theta_{1})$ and ${\bf a}_{2}(\theta_{2})$ are the array responses at the BS and the RIS-mounted drone, respectively.
We assume half-wavelength inter-element spacing to facilitate AoD and AoA estimation. The large-scale path loss for all links can be calculated as:
\begin{equation}\label{eq4}
\mathrm{Loss}[{\rm dB}] = 10\beta_1\log_{10}\!\left(\frac{r_s}{r_0}\right) + \beta_2 + 10\beta_3\log_{10}\!\left(\frac{f_s}{f_0}\right) + \beta_4,
\end{equation}
where $r_s$ and $f_s$ denote the link distance and carrier frequency, $(r_0,f_0)=(1\,\mathrm{m},1\,\mathrm{GHz})$ are reference values, $\beta_1$ and $\beta_3$ are distance and frequency attenuation, $\beta_2$ is a constant offset, and $\beta_4$ is the log-normal shadowing. This loss is consistently applied to ${\bf h}_{1,i}$, ${\bf h}_{2,i}$, and ${\bf H}$ through their large-scale gains. 

\subsection{RIS-Enabled Drone and Sensing Link}

We model the RIS reflection matrix as ${\bf \Phi}={\rm Diag}(g_1,\ldots,g_P)$ and assume equal per-element reflection power due to the UAV’s energy constraint, i.e., $|g_1|^2=\cdots=|g_P|^2=g^2$ with $0\!\le\!g\!\le\!g_{\max}$. We model the target channel as a single-bounce response
\begin{equation}\label{eq8}
{\bf T}=\beta_5\,{\bf a}_3(\theta_3){\bf a}_3^H(\theta_3)\in\mathbb{C}^{P\times P},
\end{equation}
where $\theta_3$ is the AoA at the RIS from the target, and ${\bf a}_3(\cdot)$ is the RIS array steering vector.
The channel gain $\beta_{5}$ is derived from the monostatic MIMO radar range equation given by~\cite{yu2023active}:
\begin{equation}\label{eq9}
\beta_5 = \sqrt{ (\omega^2 S) \big/ \big[(4\pi)^3 R^4\big] },
\end{equation}
where $\omega$ is the wavelength, $S$ is the target radar cross section, and $R$ represents the distance between the drone and the target.

Based on the above channel models, the two RIS-assisted reflections can be expressed as~\cite{liu2021active}:
\begin{equation}\label{eq11}
{\bf r}^{\rm refle}_1={\bf \Phi}{\bf H}{\bf s}+{\bf \Phi}{\bf n}_1,
\end{equation}
\begin{equation}\label{eq12}
{\bf r}^{\rm refle}_2={\bf \Phi}^H{\bf T}{\bf \Phi}{\bf H}{\bf s}+{\bf \Phi}^H{\bf T}{\bf \Phi}{\bf n}_1+{\bf \Phi}^H{\bf n}_2,
\end{equation}
where ${\bf n}_1$ and ${\bf n}_2$ denote the noise at the RIS input and in the echo path, respectively, following ${\bf n}_{1},{\bf n}_{2}\!\sim\!\mathcal{CN}({\bf 0},\sigma^2{\bf I})$, where $\sigma^2$ is the noise power.

\subsection{Channel-Access Attacker}

The malicious drone initiates the attack by spoofing a MAC address to gain channel access and subsequently injects additional Gaussian noise into the affected links~\cite{wei2020classification}. To remain covert, the injected noise is modeled as ${\bf n}_{\rm att}\!\sim\!\mathcal{CN}({\bf 0},\sigma_{\rm att}^2{\bf I})$, where the attacker’s noise power $\sigma_{\rm att}^2$ is upper-bounded by $\sigma^2$.

%% file: sections/section3.tex
\section{ISAC Performance Under Attack}

\subsection{Communication SINR Analysis}

With the help of the drone-mounted RIS, each user receives signals from both the direct BS-user path and the reflected BS-RIS-user path. 
Under the channel-access attack, we define the equivalent channel for user $i$ as ${\bf h}_i^{H} \triangleq {\bf h}_{1,i}^{H}{\bf \Phi}{\bf H}+{\bf h}_{2,i}^{H}$.
On this basis, the communication SINR for user $i$ is derived as:
\begin{align}\label{eq14}
&\mathrm{SINR}_i^{\mathrm{com}} = \\ \notag
&\frac{\mathbf{h}_i^{\mathrm{H}} \mathbf{R}_i \mathbf{h}_i}{\mathbf{h}_i^{\mathrm{H}} \left( \mathbf{R} - \mathbf{R}_i \right) \mathbf{h}_i + \left( \sigma^2 + \sigma_{\mathrm{att}}^2 \right) \mathbf{h}_{1,i}^{\mathrm{H}} \mathbf{\Phi} \mathbf{\Phi}^{\mathrm{H}} \mathbf{h}_{1,i} + \left( \sigma_{\mathrm{att}}^2 + \sigma_y^2 \right)},
\end{align}
where $\mathbf{R}_i$ is the covariance matrix of the signal for user~$i$ and $\sigma_y^2$ is the noise. This expression reveals how the attacker's noise power $\sigma_{\rm att}^2$ degrades performance, making the RIS's gain embedded in $\mathbf{\Phi}$ and the BS's beamforming encoded in $\mathbf{R}$ crucial defense parameters.

\subsection{Transmission of Sensing Data}

During sensing, the BS receives echo signals reflected by the drone's RIS. These echoes contain the desired reflection from the BS–drone–target–drone–BS path, the self-interference (SI) from the direct BS–drone–BS path, and thermal noise. The aggregated signal before processing can be expressed as:
\begin{equation}\label{4511111}
\mathbf{\tilde r}^{\mathrm{BS}}
= \mathbf{H}^{H}\!\left(\mathbf{r}_1^{\mathrm{refle}}+\mathbf{r}_2^{\mathrm{refle}}\right)
+ \mathbf{w}_r,
\end{equation}
where $\mathbf{w}_r\!\sim\!\mathcal{CN}(\mathbf{0},\sigma_r^2\mathbf{I}_M)$ and $\sigma^2_r$ is the noise power density.

To extract the target information, the BS performs self-interference cancellation. Considering the residual SI factor $\varepsilon$ and the additional noise injected by the channel-access attack, the processed echo becomes:
\begin{equation}\label{9265561232}
\begin{aligned}
\mathbf{r}^{\mathrm{BS}}
&= \mathbf{H}^{H}\mathbf{\Phi}^{H}\mathbf{T}\mathbf{\Phi}\mathbf{H}\mathbf{s}
 + \mathbf{H}^{H}\mathbf{\Phi}^{H}\mathbf{T}\mathbf{\Phi}\mathbf{n}'_1
 + \mathbf{H}^{H}\mathbf{\Phi}\mathbf{n}'_1 \\
&\quad + \mathbf{H}^{H}\mathbf{\Phi}^{H}\mathbf{n}'_2
 + \varepsilon\,\mathbf{H}^{H}\mathbf{\Phi}\mathbf{H}\mathbf{s}
 + \mathbf{w}'_r,
\end{aligned}
\end{equation}
where the noise distributions are now $\mathbf{n}'_1,\mathbf{n}'_2\!\sim\!\mathcal{CN}(\mathbf{0},(\sigma^2+\sigma^2_{\rm att})\mathbf{I})$ and $\mathbf{w}'_r\!\sim\!\mathcal{CN}(\mathbf{0},(\sigma_r^2+\sigma^2_{\rm att})\mathbf{I}_M)$.

The interference-plus-noise covariance matrix is given by
\begin{equation}\label{11531321}
\mathbf{J}=\mathbf{Z}+\mathbf{C}\mathbf{R}\mathbf{C}^{H},
\end{equation}
where $\mathbf{C}=\varepsilon\,\mathbf{H}^{H}\mathbf{\Phi}\mathbf{H}$ is the residual interference component. $\mathbf{Z}$ is the equivalent noise covariance matrix defined as:
\begin{align}\label{eq18}
{\bf{Z}} &= {\left( {\sigma^2  + {\sigma _{att}^2}} \right)}\left( {{{\bf{H}}^{\mathop{\rm H}\nolimits} }{\bf{\Phi }}{{\bf{\Phi }}^{\mathop{\rm H}\nolimits} }{\bf{T\Phi H}} + {{\bf{H}}^{\mathop{\rm H}\nolimits} }{{\bf{\Phi }}^{\mathop{\rm H}\nolimits} }{\bf{T\Phi }}{{\bf{\Phi }}^{\mathop{\rm H}\nolimits} }{\bf{H}}} \right) \\ \notag
&+ {\left( {\sigma^2  + {\sigma _{att}^2}} \right)}{\bf{X}}{{\bf{X}}^{\mathop{\rm H}\nolimits} }  + 2{\left( {\sigma^2  + {\sigma _{att}^2}} \right)}{{\bf{H}}^{\mathop{\rm H}\nolimits} }{{\bf{\Phi }}^{\mathop{\rm H}\nolimits} }{\bf{\Phi H}} \\ \notag
&+ {\left( {{\sigma _r^2} + {\sigma _{att}^2}} \right)}{{\bf{I}}_M},
\end{align}
with $\mathbf{X}=\mathbf{H}^{H}\mathbf{\Phi}^{H}\mathbf{T}\mathbf{\Phi}\in\mathbb{C}^{M\times P}$.

We define
\[
\mathbf{F}=\mathbf{H}^{H}\mathbf{\Phi}^{H}\mathbf{T}\mathbf{\Phi}\mathbf{H}\in\mathbb{C}^{M\times M}.
\] to represent the desired sensing signal channel. The resulting sensing SINR is the trace of the signal covariance multiplied by the inverse of the interference-plus-noise covariance:
\begin{equation}\label{48453123132}
\mathrm{SINR}^{\rm sense}=\mathrm{Tr}\!\big(\mathbf{F}\mathbf{R}\mathbf{F}^{H}\mathbf{J}^{-1}\big),
\end{equation}
where ${\rm{Tr}}\left(  \cdot  \right)$ is the trace calculator.
Finally, given the BS bandwidth $\eta$, the sensing data rate can be expressed as:
\begin{equation}\label{651231}
\gamma_{\rm sense}=\eta\log_2\!\left(1+\mathrm{SINR}^{\rm sense}\right).
\end{equation}

\subsection{Average AoI of Sensing Data}

\begin{figure}[t]
  \centering
  \includegraphics[width=7.0cm]{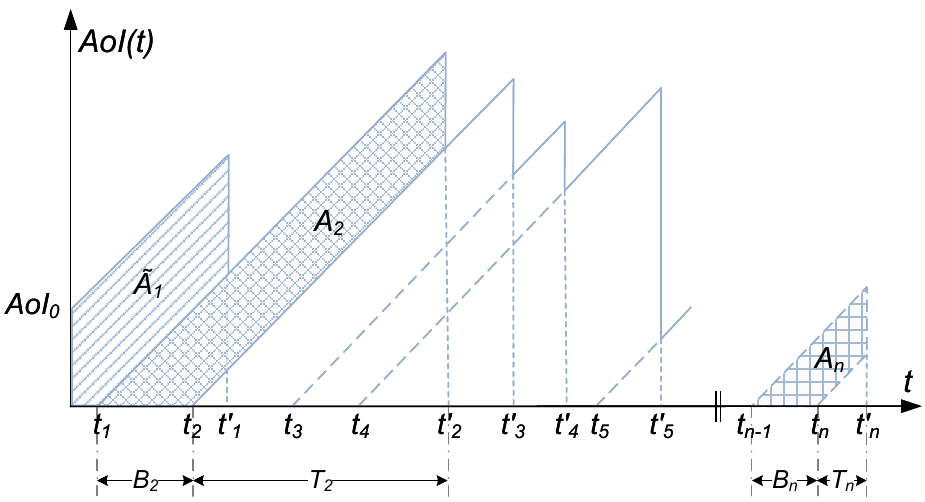}\\
  \caption{The modeling of the AoI metric.}
  \vspace{-0.3cm}
  \label{AoI}
\end{figure}

While SINR characterizes the instantaneous signal quality, it fails to capture the temporal dynamics and freshness of target information, which are essential for time-sensitive applications. To quantify this information freshness, we use the AoI metric.
As illustrated in Fig.~\ref{AoI}, the AoI evolves as a sawtooth function over time.
Let $t_i$ be the generation time of the $i$-th sensing update and $t'_i$ be its reception time at the BS. 
The total system time of this update is $T_i=t'_i-t_i$, including waiting and service durations, while the inter-arrival time between two updates is $B_i=t_i-t_{i-1}$.
Over the horizon $(0,\tau)$, the time-averaged AoI is defined as:
\begin{equation}\label{7856231335}
AAoI_\tau=\frac{1}{\tau}\int_0^\tau AoI(t)\,dt.
\end{equation}
This integral represents the area under the AoI curve, obtained by summing trapezoids between successive successful updates. The area of the $i$-th trapezoid is calculated as:
\begin{equation}\label{12545323}
{A_i} = \frac{1}{2}{\left( {{T_i} + {B_i}} \right)^2} - \frac{{T_i^2}}{2} = {B_i}{T_i} + \frac{{B_i^2}}{2}.
\end{equation}
By modeling data generation as a stochastic process and extending the observation horizon to infinity, i.e., $T\rightarrow\infty$, the long-term average AoI converges to:
\begin{equation}\label{562}
AAoI=\lambda\!\left(\mathbb{E}[B T]+\tfrac{1}{2}\mathbb{E}[B^2]\right),
\end{equation}
where $\lambda$ is the sensing-data generation rate at the BS.

To compute this metric, the sensing process is modeled as a single-server queue, with the wireless channel as the server. The service rate is equivalent to the sensing link's transmission rate, $\gamma_{\rm sense}$, and the channel utilization, representing the busy fraction, is given by~\cite{yang2023stochastic}:
\begin{equation}\label{562332}
\rho = \lambda / \gamma_{\rm sense}.
\end{equation}

This formulation provides a critical link between temporal freshness (AAoI) and the physical-layer performance ($\gamma_{\rm sense}$), establishing AoI as an explicit and optimizable metric for securing the ISAC system.

\subsection{Average AoI Model Selection}

To accurately capture the freshness of sensing data, it is necessary to choose a queuing model that reflects the characteristics of the low-altitude environment. Low-altitude applications typically involve numerous time-sensitive targets, resulting in unpredictable sensing updates and varying service times caused by changing target signals, clutter, and interference. Given this inherent randomness, we instantiate the AoI analysis using the M/M/1 queue model. This model assumes that both sensing-data arrivals and service times follow memoryless exponential distributions. Specifically, the sensing-data arrival rate is $\lambda$, and the service rate determined by channel quality is $\gamma_{\rm sense}$. For system stability and AoI convergence, the arrival rate must remain lower than the service rate.

\textbf{Theorem 1 (AAoI under M/M/1).}
When the sensing-data inter-arrival and service times are exponentially distributed with rates $\lambda$ and $\gamma_{\rm sense}$, respectively, the long-term average AoI is given by:
\begin{equation}\label{145123}
AAoI^{\rm M/M/1}
= \frac{1}{\gamma_{\rm sense}}
\left(
1+\frac{\gamma_{\rm sense}}{\lambda}
+\frac{\lambda^2}{\gamma_{\rm sense}^2-\lambda\gamma_{\rm sense}}
\right).
\end{equation}
\begin{IEEEproof} 

To derive the average AoI for the M/M/1 queue, we start from the general formula in Eq.~\eqref{562}. 
For an M/M/1 system, the packet generation follows a Poisson process, meaning the inter-arrival times $B$ are exponentially distributed with rate $\lambda$. The first two moments are therefore $\mathbb{E}[B] = 1/\lambda$ and $\mathbb{E}[B^2] = 2/\lambda^2$~\cite{kaul2012real}. 
Since the service times $O$ are also exponentially distributed with rate $\gamma_{\rm sense}$, the average service time is $\mathbb{E}\left[ {{O}} \right] = \frac{1}{{{\gamma _{sense}}}}$.
Due to the system time is $T=U+O$ (waiting time and service time), and $O$ is independent of $B$, we can write
\begin{equation}\label{956222}
\mathbb{E}\left[ {{T}{B}} \right] = \mathbb{E}\left[ {\left( {{U} + {O}} \right){B}} \right] = \mathbb{E}\left[ {{U}{B}} \right] + \mathbb{E}\left[ {{O}} \right]\mathbb{E}\left[ {{B}} \right].
\end{equation}

For the M/M/1 queue, the cross-moment between inter-arrival and waiting time can be shown to be~\cite{miller1966probability}:
\begin{equation}\label{15465562125}
\mathbb{E}\left[ {{U}{B}} \right] = \frac{\rho }{{\gamma _{sense}^2\left( {1 - \rho } \right)}}.
\end{equation}

Substituting the known expectations gives:
\begin{equation}\label{785625}
\mathbb{E}\left[ TB \right] =\frac{\rho }{{\gamma _{sense}^2\left( {1 - \rho } \right)}}+\frac{1}{\lambda \gamma _{sense}}.
\end{equation}   

Finally, we insert the expressions for $\mathbb{E}\left[ B \right]$ and $\mathbb{E}\left[ {{B^2}} \right]$ back into the general AoI formula:
\begin{equation}
    AAoI^{\rm M/M/1} = \lambda\frac{\rho }{{\gamma _{sense}^2\left( {1 - \rho } \right)}}+\frac{1}{\gamma _{sense}} + \frac{1}{\lambda}
\end{equation}
Substituting $\rho = \frac{\lambda}{\gamma _{sense}}$
and simplifying yields the expression in Theorem 1, which completes the proof.
\end{IEEEproof}

%% file: sections/section4.tex
\section{ISAC Performance Optimization for Time-Sensitive Targets under Attacks}\label{OP}

\subsection{Stackelberg Game Formulation}


We model the strategic interaction between the malicious attacker and the LAWN defenders as a three-player Stackelberg game.
In this hierarchical game, the malicious drone serves as the leader because it initiates the channel access attack before any defensive measures can be taken, disrupting the network and compelling the defenders to react.
Afterward, the legitimate RIS-equipped drone and the ISAC BS act as the second and first followers, sequentially adjusting their reflection gain and sensing data generation rate to counter the attacker’s impact.
The objective of this hierarchical game is to derive a stable Stackelberg equilibrium where all players’ utilities converge, representing the balance between the attacker’s aggressiveness and the defenders’ adaptive countermeasures.

Each player’s utility depends on two core performance metrics: the AAoI and the average communication SINR for all users (ASINR), defined as:
\begin{equation}\label{521455}
{\mathop{\rm ASINR}\nolimits}  = \frac{1}{N}\sum\nolimits_{i = 1}^N {{\mathop{\rm SINR}\nolimits} _i^{com}}.
\end{equation}

\subsection{Player Sub-Games}
Each player’s utility is modeled as a weighted combination of the two performance metrics, where $\zeta_1, \zeta_2 > 0$ balance the trade-off between sensing freshness and communication quality.
Additionally, each player’s action incurs a corresponding cost, modeled as follows:

1. \textbf{First Follower (BS):} The BS aims to minimize the AAoI while accounting for its data-generation cost ${\vartheta _{BS}}$. It optimizes the sensing data generation rate $\lambda$, and its utility is given by
    \begin{equation}\label{596211548696221}
{\varphi _{BS}} =  - {\zeta _1}AAoI + {\zeta _2}{\mathop{\rm ASINR}\nolimits}  - {\vartheta _{BS}}\lambda  ,
\end{equation}

2. \textbf{Second Follower (RIS Drone):} The legitimate drone minimizes the AAoI while considering its power amplification cost ($\vartheta _{RIS}$). It optimizes the RIS gain $g$, and its utility function is expressed as
\begin{equation}\label{7854785}
{\varphi _{RIS}} =  - {\zeta _1}AAoI + {\zeta _2}{\mathop{\rm ASINR}\nolimits}  - {\vartheta _{RIS}}g,
\end{equation}

3. \textbf{Leader (Attacker):} The attacker aims to degrade the ISAC performance by increasing the AoI and reducing the ASINR, subject to its attack cost ${\vartheta_{att}}$.
It optimizes the added noise power $\sigma_{att}^2$, and its utility is defined as
  \begin{equation}\label{44562}
{\varphi _{att}} = {\zeta _1}AAoI - {\zeta _2}{\mathop{\rm ASINR}\nolimits}  - {\vartheta _{att}}{\sigma _{att}},
\end{equation}

\subsection{Equilibrium Solution via Backward Induction}

The overall problem can be reformulated as a three-layer nested optimization process that is solved sequentially from the inner (first follower) to the outer (leader) layer:
\begin{equation}\label{51155}
\begin{cases}
\begin{aligned}
& \underset{\sigma_{\text{att}}}{\text{max}} \; \varphi_{\text{att}}\left( {\lambda ,g,{\sigma _{\text{att}}}} \right) \\
& \text{s.t.} \;\quad 0 \le \sigma_{\text{att}} \le \sigma \\
& \text{SINR}_i^{\text{com}} < \text{SINR}_{\text{thre}} \\
& \text{Optimal solution:} \; \left( {{\lambda _{\text{opt}}},{g_{\text{opt}}},{\sigma _{\text{att}}}} \right)
\end{aligned} \\
\begin{cases}
\begin{aligned}
& \underset{g}{\text{max}} \; \varphi_{\text{RIS}}\left( {\lambda ,g,{\sigma _{\text{att}}}} \right) \\
& \text{s.t.} \;\quad 0 \le g \le {g_{\max }} \\
& \text{SINR}_i^{\text{com}} \ge \text{SINR}_{\text{thre}} \\
& \text{Optimal solution:} \; \left( {{\lambda _{\text{opt}}},g,{\sigma _{\text{att}}}} \right)
\end{aligned} \\
\begin{cases}
\begin{aligned}
& \underset{\lambda}{\text{max}} \; \varphi_{\text{BS}}\left( {\lambda ,g,{\sigma _{\text{att}}}} \right) \\
& \text{s.t.} \;\quad 0 \le \lambda \le {\gamma _{\text{radar}}} \\
& \text{SINR}_i^{\text{com}} \ge \text{SINR}_{\text{thre}}
\end{aligned}
\end{cases}
\end{cases}
\end{cases},
\end{equation}
where ${{\lambda _{opt}}}$ and ${{g_{opt}}}$ represent optimized values for variable $\lambda$ and $g$, respectively.
This sequential optimization ensures that the leader’s optimal attacking strategy is determined with full anticipation of the followers’ best responses, guaranteeing the existence of a Stackelberg equilibrium.

To solve the formulated three-layer Stackelberg game, we develop a two-level backward-induction (BI) framework. At the lower level, each player’s sub-problem is solved individually.
Because the utility functions are complex and lack closed-form solutions, a numerical, derivative-free optimization method is employed.
Specifically, the Golden Section Search and Parabolic Interpolation (GSSPI) method is adopted~\cite{yang2024can}.
This method is well-suited for non-differentiable objectives, efficiently obtaining each player’s optimal action by combining the guaranteed convergence of golden-section search with the quadratic convergence speed of parabolic interpolation.

At the higher level, an iterative BI algorithm is executed to obtain the overall Stackelberg equilibrium.
Following the hierarchical game logic, the optimization proceeds from the last mover to the first in each iteration.
First, given the current strategies of the other players, the algorithm solves the BS’s sub-game using GSSPI to determine its optimal sensing rate $\lambda$.
Next, it solves the RIS drone’s sub-game to obtain the optimal amplification gain $g$, in response to the attacker’s strategy and the BS’s updated decision.
Finally, the algorithm addresses the attacker’s sub-game to derive the optimal attack power $\sigma_{att}$ that maximizes its utility while anticipating the followers’ optimal responses.

This iterative process continues until the strategies and utilities of all three players converge, thereby yielding the Stackelberg equilibrium solutions.

%% file: sections/section5.tex
\section{Evaluations and Analysis}\label{EA}


\begin{figure*}[!htb]
\centering
\subfigure[]
{
    \label{tu1}
    \includegraphics[height=0.18\textwidth]{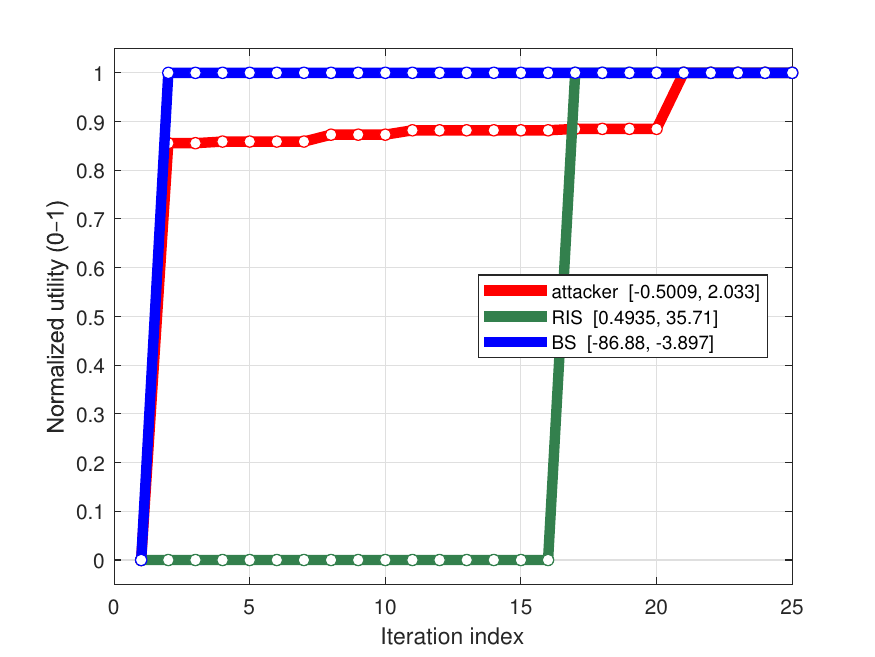}
}
\quad
\hspace{-0.3in}
\subfigure[]
{
   \label{tu2}
    \includegraphics[height=0.18\textwidth]{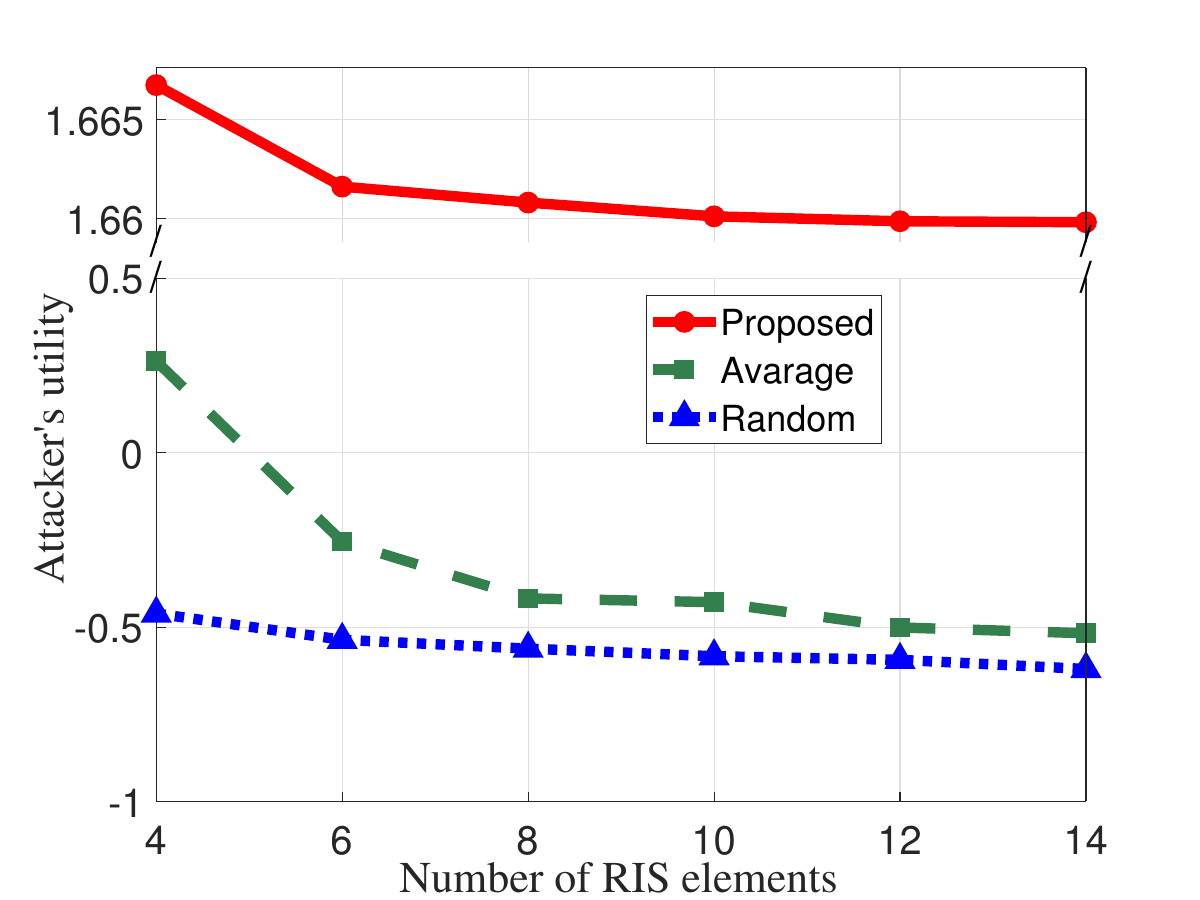}
}
\quad
\hspace{-0.3in}
\subfigure[]
{
   \label{tu3}
    \raisebox{-1.2mm}{\includegraphics[height=0.18\textwidth]{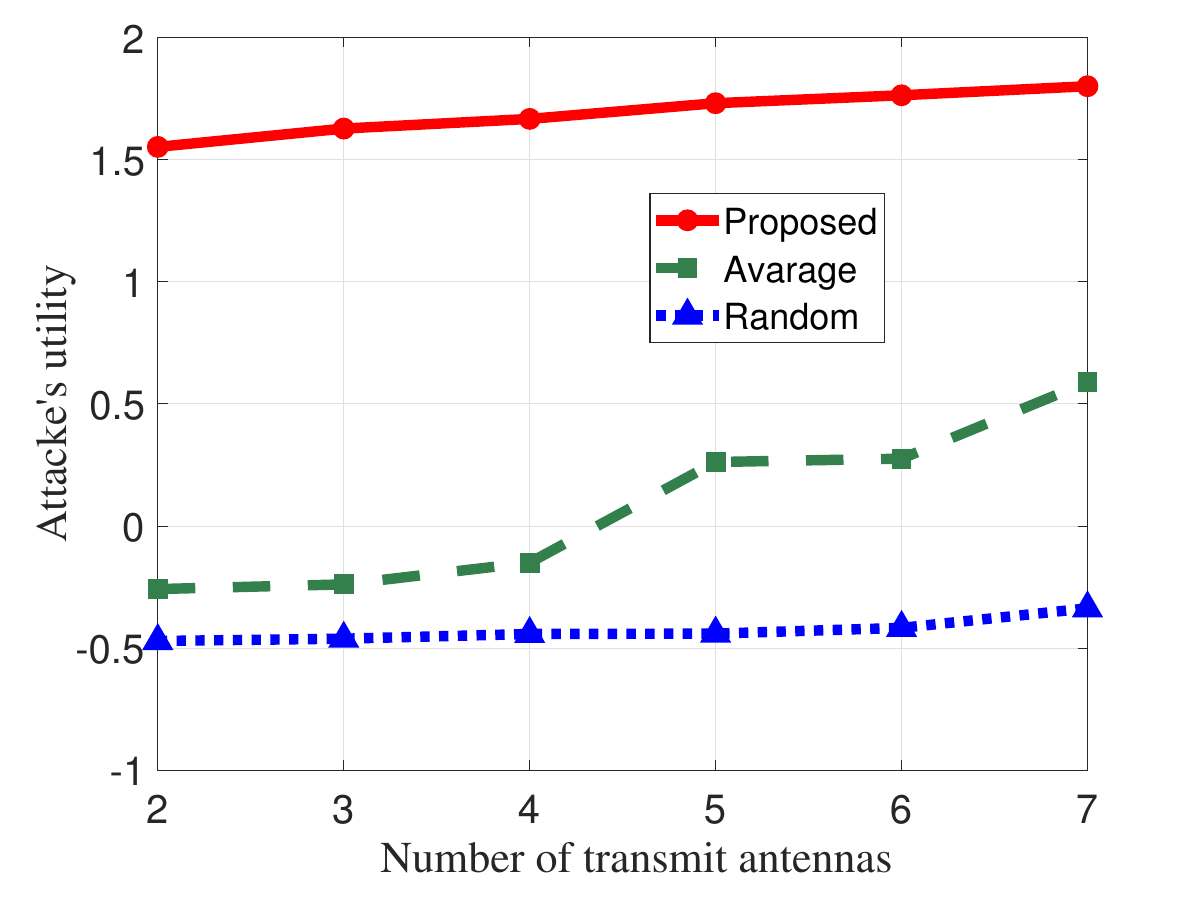}}
}
\quad
\hspace{-0.3in}
\subfigure[]
{
   \label{tu4}
    \raisebox{-0.2mm}{\includegraphics[height=0.18\textwidth]{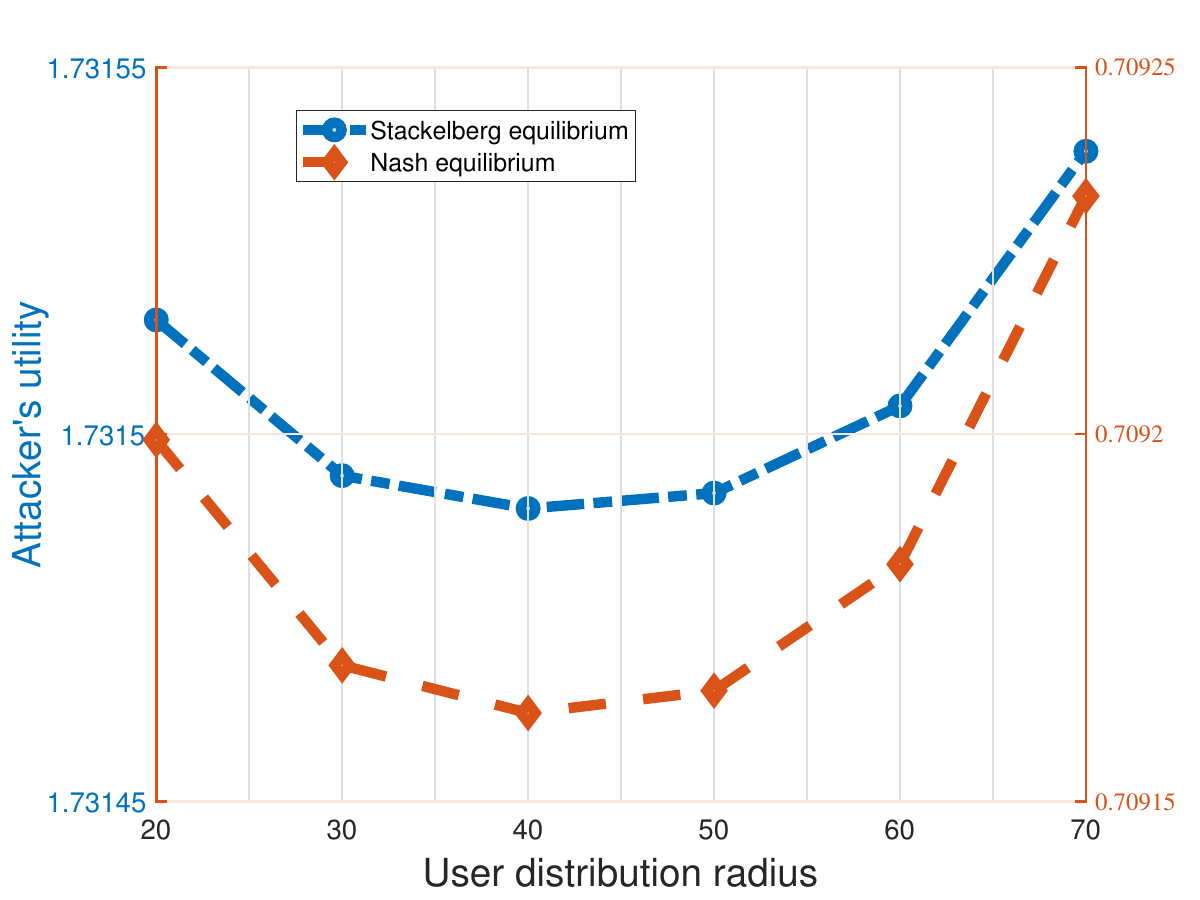}}
}
\quad
\hspace{-0.3in}
\caption{
Convergence and performance analysis of the proposed Stackelberg game-based defense algorithm.
(a) Convergence behavior of the utilities of all players.
(b) Attacker’s utility versus the number of RIS elements.
(c) Attacker’s utility versus the number of transmit antennas.
(d) Attacker’s utility versus the user distribution radius with different equilibriums.}

\label{tu123}
\end{figure*}

\begin{figure*}[!htb]
\centering
\subfigure[ ]
{
    \label{tuadd1}
    \includegraphics[width=0.60\columnwidth]{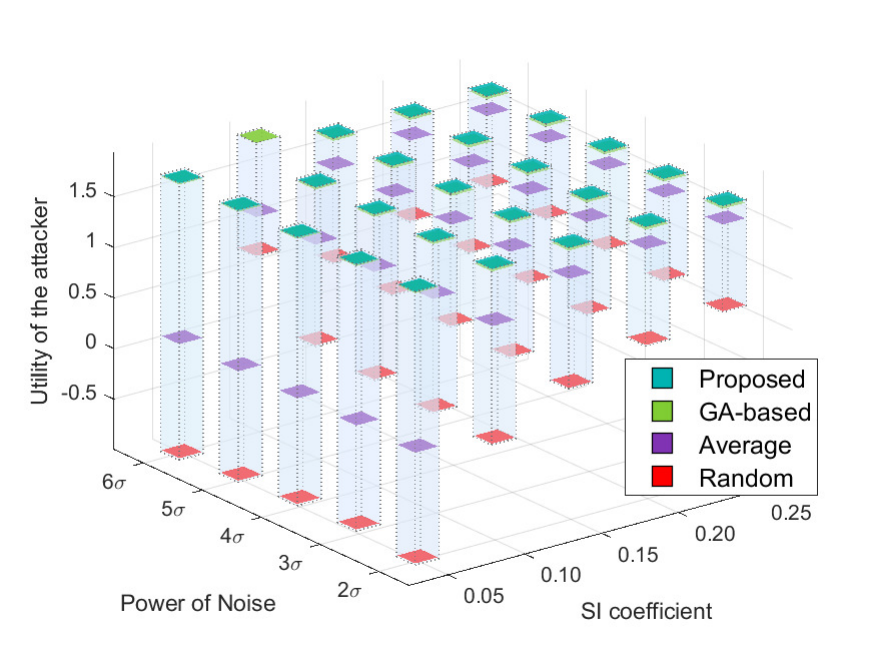}
}
\qquad
\hspace{-0.3in}
\subfigure[ ]
{
   \label{tuadd2}
    \includegraphics[width=0.60\columnwidth]{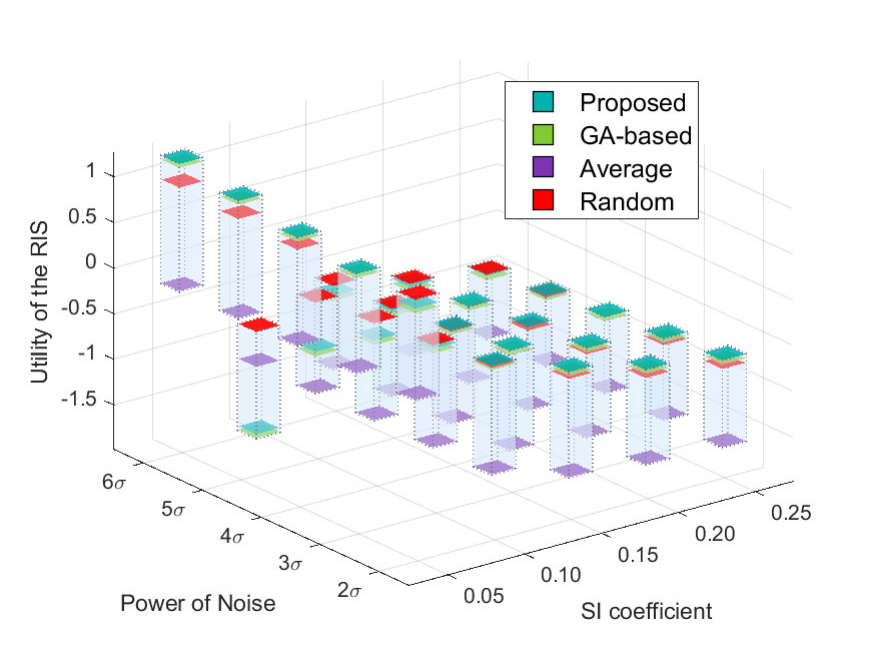}
}
\qquad
\hspace{-0.3in}
\subfigure[ ]
{
   \label{tuadd3}
    \includegraphics[width=0.60\columnwidth]{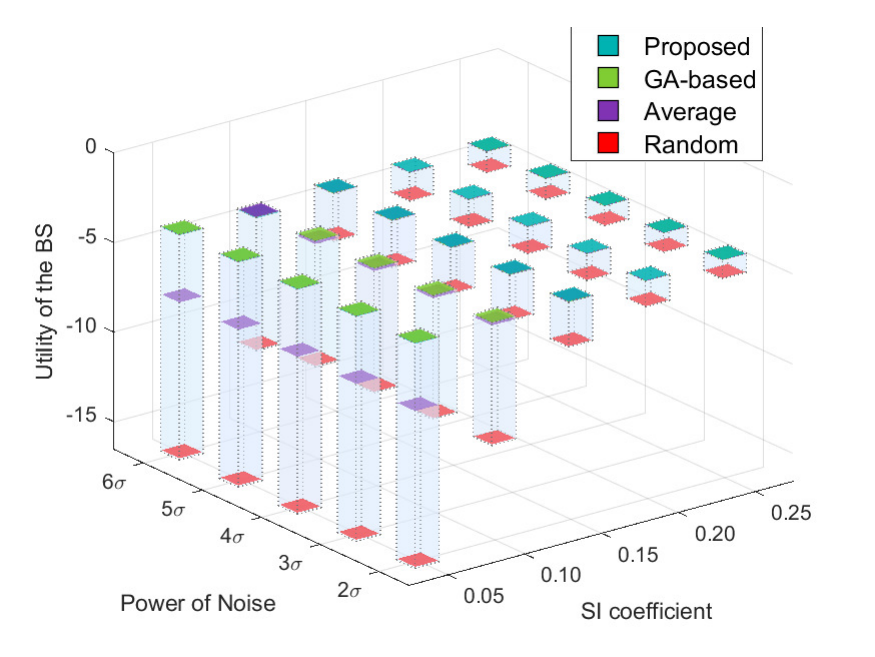}
}
\quad
\hspace{-0.3in}
\caption{Performance comparison of the proposed and GA-based optimization schemes. (a) The attacker's utility with different SI coefficients and attacking noise. (b) The RIS's utility with different SI coefficients and attacking noise. (c) The BS's utility with different SI coefficients and attacking noise.}
\vspace{-0.3cm}
\label{tuadd}
\end{figure*}

This section evaluates the proposed algorithm under multiple configurations.
We first analyze its convergence behavior in terms of utilities, optimized variables, and performance metrics.
Then, we compare its performance against three benchmark schemes: average allocation, random allocation, and Genetic Algorithm (GA)-based optimization.

\subsection{Parameter Setting}

In the considered low-altitude network, a multi-antenna BS communicates with two single-antenna users and senses one target with the assistance of an RIS-equipped legitimate drone.
The key simulation parameters are summarized in Table~\ref{parameters}, and the SINR threshold is fixed at 5 dB.

For comparison, three benchmark methods are considered:
the average scheme, which uniformly allocates resources across the feasible domain to ensure fairness but ignores channel variations~\cite{chang2021resource};
the random scheme, which randomly samples parameters from uniform distributions as a non-intelligent baseline~\cite{birhanie2020stochastic};
and the GA-based optimization, which iteratively evolves candidate solutions through selection, crossover, and mutation to achieve near-optimal performance at a high computational cost~\cite{ma2024optimizing}.

\subsection{Convergence and Performance Analysis}

The convergence behavior of the proposed Stackelberg game-based defense algorithm is illustrated in Fig.~\ref{tu1}. As shown, the utilities of all three players (the attacker, RIS-equipped drone, and BS) gradually stabilize, confirming that the BI process converges to a Stackelberg equilibrium. Each player’s utility increases during the iterations, as the non-cooperative game encourages every participant to selfishly and rationally maximize their own benefit. At equilibrium, the attacker, RIS drone, and BS reach stable utilities of 2.033, 35.71, and –3.897, respectively, validating the convergence and rationality of the proposed framework.

Figs.~\ref{tu2}–\ref{tu3} further depict the attacker’s utility under varying system parameters. The proposed method achieves the highest utility (1.659), outperforming the average (–0.517) and random (–0.62) allocation schemes. As the number of RIS elements increases, the attacker’s utility slightly declines because stronger reflections enhance SINR and suppress interference. Conversely, more BS transmit antennas expand the spatial attack surface, increasing the attacker’s utility. 
The fact that our model finds an equilibrium with a higher attacker utility demonstrates its effectiveness in modeling the strategic conflict and balancing the objectives among all players.

As demonstrated in Fig.~\ref{tuadd}, the proposed method consistently outperforms the GA-based, Average, and Random baselines across all players' utilities. This superior performance stems from the GSSPI-based BI algorithm, which leverages the hierarchical game structure to find a more efficient and superior equilibrium. Furthermore, the results show that as the attacker’s noise power increases, the utilities of the attacker and the RIS drone also increase, whereas the BS utility remains stable. This dynamic reflects the system’s adaptive balance, where defenders adjust their strategies to mitigate stronger attacks effectively.

Finally, Fig.~\ref{tu4} contrasts the Stackelberg and Nash equilibria. The Stackelberg framework yields higher attacker utility by capturing sequential causality among players, whereas the static Nash model overlooks hierarchical interactions, leading to inferior outcomes. As the user-distribution radius expands from 20~m to 40~m, the average SINR improves and the attacker’s utility decreases. Moreover, beyond 50~m, the SINR declines, reversing the trend. In summary, the proposed algorithm demonstrates faster convergence, higher equilibrium utilities, and more robust ISAC performance under adversarial conditions.

\begin{table}[]
\centering
\caption{Simulation settings.}
\label{parameters}
\resizebox{0.45\textwidth}{!}{
\begin{tabular}{c c c c}
\toprule
\textbf{Symbol} & \textbf{Value} & \textbf{Symbol} & \textbf{Value} \\
\midrule
$M$ & 4 & $\sigma^2$ & -174 dBm/Hz \\
$N$ & 2 & $P$ & 16 \\
$(x_{BS}, y_{BS}, z_{BS})$ & $[0, 0, 1.5]$ & $(x_{RIS}, y_{RIS}, z_{RIS})$ & $[5, 10, 1.5]$ \\
$(x_{user}, y_{user}, z_{user})$ & $[30, 10, 1.5]$ & $(x_{tar}, y_{tar}, z_{tar})$ & $[0, 60, 1.5]$ \\
$r_{user}$ & 10 & $\varepsilon $ & 0.1 \\
$\eta$ & $10^7$ Hz & $P_{trans}$ & 0.9 \\
$\sigma^2_{y}$ & -174 dBm/Hz & $\sigma^2_{r}$ & -174 dBm/Hz \\
$g_{max}$ & 1 & $Iter_{max}$ & 25 \\
\bottomrule
\end{tabular}
}
\vspace{-0.3cm}
\end{table}

%% file: ref.bib
@article{zhao2025generative,
  title={Generative {AI}-enabled wireless communications for robust low-altitude economy networking},
  author={Zhao, Changyuan and Wang, Jiacheng and Zhang, Ruichen and Niyato, Dusit and Sun, Geng and Du, Hongyang and Kim, Dong In and Jamalipour, Abbas},
  journal={IEEE Wireless Commun.},
  year={2025}
}

@article{jiang2025integrated,
  title={Integrated sensing and communication for low altitude economy: Opportunities and challenges},
  author={Jiang, Yihang and Li, Xiaoyang and Zhu, Guangxu and Li, Hang and Deng, Jing and Han, Kaifeng and Shen, Chao and Shi, Qingjiang and Zhang, Rui},
  journal={IEEE Commun. Mag.},
  year={2025},
  publisher={IEEE}
}

@article{wang2024generative,
  title={Generative {AI} for integrated sensing and communication: Insights from the physical layer perspective},
  author={Wang, Jiacheng and Du, Hongyang and Niyato, Dusit and Kang, Jiawen and Cui, Shuguang and Shen, Xuemin and Zhang, Ping},
  journal={IEEE Trans. Wireless Commun.},
  volume={31},
  number={5},
  pages={246--255},
  year={2024},
  publisher={IEEE}
}

@article{liu2024game,
  title={A game theoretical anti-jamming beamforming approach for integrated sensing and communications systems},
  author={Liu, Yuan and Zhang, Bangning and Guo, Daoxing and Wang, Haichao and Ding, Guoru and Yang, Ning and Gu, Jiangchun},
  journal={IEEE Trans. Veh. Technol.},
  year={2024},
  publisher={IEEE}
}

@article{liu2020joint,
  title={Joint transmit beamforming for multiuser MIMO communications and MIMO radar},
  author={Liu, Xiang and Huang, Tianyao and Shlezinger, Nir and Liu, Yimin and Zhou, Jie and Eldar, Yonina C},
  journal={IEEE Transactions on Signal Processing},
  volume={68},
  pages={3929--3944},
  year={2020},
  publisher={IEEE}
}

@article{yu2023active,
  title={Active {RIS}-aided {ISAC} systems: Beamforming design and performance analysis},
  author={Yu, Zhiyuan and Ren, Hong and Pan, Cunhua and Zhou, Gui and Wang, Boshi and Dong, Mianxiong and Wang, Jiangzhou},
  journal={IEEE Trans. Commun.},
  volume={72},
  number={3},
  pages={1578--1595},
  year={2023},
  publisher={IEEE}
}

@article{liu2021active,
  title={Active reconfigurable intelligent surface: Fully-connected or sub-connected?},
  author={Liu, Kunzan and Zhang, Zijian and Dai, Linglong and Xu, Shenheng and Yang, Fan},
  journal={IEEE Commun. Letters},
  volume={26},
  number={1},
  pages={167--171},
  year={2021},
  publisher={IEEE}
}

@inproceedings{wei2020classification,
  author    = {Xianglin Wei and Li Li and Chaogang Tang and Milo{\v{s}} Doroslova{\v{c}}ki and Suresh Subramaniam},
  title     = {Classification of Channel Access Attacks in Wireless Networks: A Deep Learning Approach},
  booktitle = {Proc. {IEEE} 40th Int. Conf. Distrib. Comput. Syst. ({ICDCS})},
  pages     = {809--819},
  year      = {2020}
}

@article{yang2024can,
  title={Can We Realize Data Freshness Optimization for Privacy Preserving-Mobile Crowdsensing With Artificial Noise?},
  author={Yang, Yaoqi and Zhang, Bangning and Guo, Daoxing and Xiong, Zehui and Niyato, Dusit and Han, Zhu},
  journal={IEEE Trans. Mob. Comput.},
  year={2024},
  publisher={IEEE}
}

@inproceedings{kaul2012real,
  author    = {Sanjit Kaul and Roy Yates and Marco Gruteser},
  title     = {Real-Time Status: How Often Should One Update?},
  booktitle = {Proc. {IEEE} 31st Int. Conf. Comput. Commun. ({INFOCOM})},
  pages     = {2731--2735},
  year      = {2012}
}

@article{yang2023stochastic,
  title={Stochastic geometry-based age of information performance analysis for privacy preservation-oriented mobile crowdsensing},
  author={Yang, Yaoqi and Zhang, Bangning and Guo, Daoxing and Wang, Weizheng and Nie, Jiangtian and Xiong, Zehui and Xu, Renhui and Zhou, Xiaokang},
  journal={IEEE Trans. Veh. Technol.},
  volume={72},
  number={7},
  pages={9527--9541},
  year={2023},
  publisher={IEEE}
}

@misc{miller1966probability,
  title={Probability, random variables, and stochastic processes},
  author={Miller, Irwin},
  year={1966},
  publisher={Taylor \& Francis}
}

@article{ma2024optimizing,
  title={Optimizing joint technology selection, production planning and pricing decisions under emission tax: A Stackelberg game model and nested genetic algorithm},
  author={Ma, Shuang and Zhang, Linda L and Cai, Xiaotian},
  journal={Expert Systems with Applications},
  volume={238},
  pages={122085},
  year={2024},
  publisher={Elsevier}
}

@article{chang2021resource,
  title={Resource-allocation mechanism: Game-theory analysis},
  author={Chang, Shu-Lin and Lee, Kun-Chang and Huang, Ruey-Rong and Liao, Yu-Hsien},
  journal={Symmetry},
  volume={13},
  number={5},
  pages={799},
  year={2021},
  publisher={MDPI}
}

@inproceedings{birhanie2020stochastic,
  author    = {Habtamu Mohammed Birhanie and Sidi-Mohammed Senouc and Mohammed Ayoub Messous and Amel Arfaoui and Ali Kies},
  title     = {A Stochastic Theoretical Game Approach for Resource Allocation in Vehicular Fog Computing},
  booktitle = {Proc. {IEEE} 17th Annu. Consum. Commun. Netw. Conf. ({CCNC})},
  pages     = {1--2},
  year      = {2020}
}
